# Melting of Charge Stripes in Vibrationally Driven $La_{1.875}Ba_{0.125}CuO_4$: Assessing the Respective Roles of Electronic and Lattice Order in Frustrated Superconductors

M. Först[1*], R.I. Tobey[2], H. Bromberger[1], S.B. Wilkins[3], V. Khanna[4,5], A.D. Caviglia[1], Y.-D. Chuang[6], W.S. Lee[7], W.F. Schlotter[8], J.J. Turner[8], M.P. Minitti[8], O. Krupin[9], Z.J. Xu[3], J.S. Wen[3], G.D. Gu[3], S.S. Dhesi[5], A. Cavalleri[1,4†], and J.P. Hill[3§]

[1]*Max-Planck Institute for the Structure and Dynamics of Matter, Hamburg, Germany*
[2]*Zernike Institute for Advanced Materials, University of Groningen, Groningen, The Netherlands*
[3]*Condensed Matter Physics and Materials Science Department, Brookhaven National Laboratory, Upton, NY*
[4]*Department of Physics, Clarendon Laboratory, University of Oxford, United Kingdom*
[5]*Diamond Light Source, Chilton, Didcot, Oxfordshire, United Kingdom*
[6]*Advanced Light Source, Lawrence Berkeley Laboratory, Berkeley, CA*
[7]*SIMES, SLAC National Accelerator Laboratory and Stanford University, Menlo Park, CA*
[8]*Linac Coherent Light Source, SLAC National Accelerator Laboratory, Menlo Park, CA*
[9]*European XFEL GmbH, Hamburg, Germany*

## Abstract

We report femtosecond resonant soft X-ray diffraction measurements of the dynamics of the charge order and of the crystal lattice in non-superconducting, stripe-ordered $La_{1.875}Ba_{0.125}CuO_4$. Excitation of the in-plane Cu-O stretching phonon with a mid-infrared pulse has been previously shown to induce a transient superconducting state in the closely related compound $La_{1.675}Eu_{0.2}Sr_{0.125}CuO_4$. In $La_{1.875}Ba_{0.125}CuO_4$, we find that the charge stripe order melts promptly on a sub-picosecond time scale. Surprisingly, the low temperature tetragonal distortion is only weakly reduced, reacting on significantly longer time scales that do not correlate with light-induced superconductivity. This experiment suggests that charge modulations alone, and not the LTT distortion, prevent superconductivity in equilibrium.

Cuprate compounds of the type La$_{2-x}$Ba$_x$CuO$_4$ exhibit bulk superconductivity for doping levels between $x$=0.05 and $x$=0.25. The maximum critical temperature, $T_C$=30 K, occurs for $x$=0.16. Importantly, the superconducting phase is strongly suppressed in a narrow doping range around $x$=1/8 [1,2], where the so-called "stripe phase" [3,4] is found. In this phase, the doped holes form chains within the CuO$_2$ planes, which separate regions of oppositely phased antiferromagnetism [4]. Co-existing with this "stripe" order is a low-temperature tetragonal (LTT) distortion of the crystal lattice, setting in at the same temperature [4].

The relationship between the superconducting state, stripe order, and the LTT distortion is one of the great mysteries in high-$T_C$ superconductivity. It has long been believed that stripes are pinned by the LTT distortion [5,6] and compete with superconductivity to result in an incoherent state with non-superconducting transport. However, it has become clear that the situation is considerably more subtle. Recent experiments have suggested that stripe order can also exist without the LTT distortion, as observed in a low-temperature high-pressure phase [7]. Furthermore, there is evidence that stripes are compatible with in-plane Cooper pairing, creating phases that may involve 2D superconducting CuO$_2$ planes de-coupled as a result of the periodic charge modulation, which, it is speculated, may prevent interlayer Josephson tunneling [8,9].

Recently, the richness of the behavior of these stripe-ordered cuprates has been further emphasized by Fausti *et al.* with the discovery of a light-induced superconducting state in the non-superconducting 1/8-doped La$_{1.675}$Eu$_{0.2}$Sr$_{0.125}$CuO$_4$ (LESCO$_{1/8}$) [10]. Like La$_{1.875}$Ba$_{0.125}$CuO$_4$ (LBCO$_{1/8}$), LESCO$_{1/8}$ exhibits an LTT phase (below 135 K) and charge stripe order (below 25 K) [11]. Fausti *et al.* used a femtosecond mid-infrared pulse to resonantly excite a 600 cm$^{-1}$ infrared-active, in-plane Cu-O stretching mode. Transient superconductivity along the *c*-axis was evidenced by the appearance of a Josephson plasma edge in the THz optical response, setting in within the 1-2 ps time resolution of the THz probe. The authors hypothesized that the light-induced Josephson coupling of the $CuO_2$ planes is a consequence of an instantaneous stripe order melting, mediated by the direct distortion of the lattice. The fast time scale of this re-coupling was taken as indirect support for the picture of a 2D "pair-density wave" coherent state, as discussed in Ref. 8, in which the stripe order suppresses superconductivity only along the c-axis.

However, the fate of the charge stripe order and of the LTT distortion in the transient 3D coherent superconductor following the mid-IR excitation pulse was not probed in this experiment and remains unknown. Knowledge of their dynamics would provide new insight into the microscopic physics of this light-induced superconductivity, and perhaps even into the origin of superconductivity itself. In particular, by understanding which order needs to be de-stabilized for superconductivity to appear, one would be able to assess ´cause and effect´ relationships in this complex system where several types of order compete at equilibrium. Such assessment, however, can only be made by examining the ultrafast response following controlled excitation out of the equilibrium state.

Here, we use femtosecond resonant soft X-ray diffraction at a free electron laser [12,13,14,15] to directly probe the dynamics of both the stripe order and the LTT distortion in the stripe-ordered cuprate $La_{1.875}Ba_{0.125}CuO_4$ ($LBCO_{1/8}$) following a mid-IR pump. We find that excitation in resonance with the same Cu-O stretch mode used to induce superconductivity in $LESCO_{1/8}$ causes a non-equilibrium melting of the charge stripe order. Importantly, the stripe order suppression, which is prompt and almost complete, is decoupled from the LTT distortion. This latter is only reduced by 12% and occurs over timescales that are significantly longer than both the charge order melting, and the light-induced superconducting transition in $LESCO_{1/8}$. This measurement unveils a direct correlation between charge stripe order and the frustration of superconductivity in the 214 cuprates at 1/8 doping, and suggests that the lattice distortion may simply be an epiphenomenon.

For LBCO$_{1/8}$, in thermal equilibrium, one-dimensional hole ordering within the CuO$_2$ planes is observed below $T_{CO}$=55 K. As sketched in Fig. 1(a), the charges localize in stripes that form π-phase shift domain walls for the antiferromagnetism. These stripes are rotated by 90° in neighboring planes along the crystal *c*-axis [9]. In addition, parallel stripes are shifted by half a period between every second CuO$_2$ plane, giving rise to charge order reflections at a wave vector of (0.24 0 0.5). Here, we use the conventional notation of the high temperature tetragonal (I4/*mmm*) unit cell. At the same temperature $T_{LTT} = T_{CO}$, the LBCO$_{1/8}$ crystal lattice transforms from the low-temperature orthorhombic phase into the LTT phase, where the CuO$_6$ octahedra tilt around the [100] and [010] axes, i.e. along the O-Cu-O bonds, in the *ab*-plane. This distortion buckles the CuO$_2$ planes changing tilt direction from layer to layer [4]. The critical temperature $T_C$ for bulk superconductivity in LBCO$_{1/8}$ is <3 K [1,2,7].

Both static stripe order and the LTT distortion can be measured through resonant soft X-ray diffraction near the oxygen K-edge. Static charge stripes are observed at a wave vector of $Q$ = (0.24 0 0.5) with a large resonant enhancement at the mobile carrier peak [16,17,18]. This is a pre-edge feature in the X-ray absorption spectrum at 528 eV, associated with the doped holes [19]. The LTT distortion can be directly measured through the (001) diffraction peak which is structurally forbidden in the high-temperature phases. In the low-temperature phase, it becomes allowed, on resonance, because of the rotation of the sense of tilt in adjacent CuO$_2$ planes [20]. This peak can be observed by tuning the photon energy to 532 eV [17,18], corresponding to the La-O hybridized states. We note that we have chosen to study LBCO$_{1/8}$ and not LESCO$_{1/8}$ in this work because the stripe order peak intensity is approximately ten times larger than in LESCO$_{1/8}$ and the present experiments are extremely challenging at current X-ray free electron lasers. They would not be possible in LESCO$_{1/8}$. Nevertheless, the comparison between stripe order melting in LBCO$_{1/8}$ and light-induced superconductivity in LESCO$_{1/8}$ is expected to be a valid one. Both compounds exhibit the same low temperature ground state, with both an LTT distortion and charge and spin stripe order. Both compounds are non-superconductors at 1/8 doping and both are equilibrium superconductors with a maximum T$_C$ ~ 20-30 K. Thus, this work is expected to shed light on the microscopic mechanisms for photo-induced superconductivity in stripe-ordered cuprates and possibly on the respective roles of electronic and lattice order in suppressing equilibrium superconductivity at x=1/8 doping.

Time-resolved resonant soft X-ray diffraction experiments were performed at the SXR beamline of the Linac Coherent Light Source (LCLS) [21,22]. A schematic drawing of the experimental setup is shown in Figure 1(b). The LBCO$_{1/8}$ single crystal, grown by the floating zone technique, was cleaved to reveal the (001) surface for these measurements. The sample was held at base temperature T=13 K in the stripe-ordered, LTT-distorted phase. Femtosecond mid-infrared pulses, derived from an optical parametric amplifier and subsequent difference frequency mixing, were used for excitation. The central photon energy was tuned to the 85 meV (14.5 μm wavelength) resonance of the infrared-active in-plane Cu-O stretching vibration [23]. The excitation pulses were 200 fs long, polarized in the *ab*-plane, and focused onto the sample with a spot size of 700 μm. In our experiments, the excitation fluence was kept at 1.9 mJ/cm$^2$, equal to the conditions of the light-induced superconductivity transition studied in LESCO$_{1/8}$ [10].

Synchronized X-ray pulses of sub-100 fs duration, tuned to photon energies of 528 and 532 eV for the (0.24 0 0.5) and (001) wave vectors, respectively, were selected by a grating monochromator providing a bandwidth of approximately 1.5 eV. The X-ray beam was aligned collinearly with the mid-IR pump and focused onto the sample with a 200 μm diameter. The absorption length of the X-ray pulses around the oxygen K-edge is approximately 200 nm. Together with the 1-2 μm extinction depth of the mid-IR light at the phonon resonance, this means that the x-rays are probing a homogenously pumped sample volume. A high-vacuum diffraction chamber, equipped with a fast-readout CCD camera, was used for the experiments [24]. The measurements were performed with a 60 Hz repetition rate. The time resolution of this experiment was 300 fs, limited by the timing jitter between the synchronized X-ray and optical laser pulses.

The time-dependent response of the (0.24 0 0.5) stripe order diffraction peak to the optical excitation is shown in Figure 2. The upper panel shows the diffracted X-ray beam, recorded with the CCD camera and averaged over about 20,000 FEL shots. The vertical lines in the images result from the detector as a result of inherent gain and offset variations amongst the different active arrays. At negative time delays, a broad and rather weak peak is observed on the CCD camera, consistent with the short correlation length of the stripe order in $LBCO_{1/8}$ [16,17]. The lower panel shows the integrated transient intensity of this diffraction peak. These data points were obtained by dark image correction, a background subtraction and pixel integration within the region of interest. The integrated diffraction intensity of this peak promptly decreases by about 70% after the arrival of the mid-IR excitation. These results show that stripe order is melted on a sub-picosecond timescale by these mid-IR pulses. The red solid line represents a single-exponential function used to visualize the reduction of the scattering intensity, with a time constant set to the 300 fs time resolution of the experiment. The reduction of integrated scattered intensity is also apparent in the camera images, e.g., at a time delay of +1.3 ps shown on the top right. The fast timescale observed here is similar to the one observed in the $LESCO_{1/8}$ THz probe experiments, implying that the ultrafast formation of the superconducting state and the melting of charge modulations are connected.

In contrast, the evolution of the LTT phase, as probed by the (001) diffraction peak, is very different from that of the stripe order. CCD images of this diffraction peak, averaged over 400 FEL shots taken at a positive and a negative time delay, are shown in Fig. 3 together with the transient integrated intensity. The integrated scattered intensity of this structural – and therefore sharp and intense – peak drops by only 12%, and on a much longer time scale. A single-exponential decay fitted to the data (red solid line) yields a time constant of 15 ps. This timescale is likely set by acoustic propagation, as the relaxation of the LTT distortion requires the lattice planes to expand, a process that is limited by the speed of sound.

Our experiments demonstrate that the resonant mid-IR excitation in $LBCO_{1/8}$ triggers the ultrafast formation of a non-equilibrium state in which stripe charge correlations have disappeared while the LTT distortion still exists. This decoupling is not present in the equilibrium phase diagram of LBCO. Given the prompt appearance of 3D superconductivity in $LESCO_{1/8}$ under identical conditions, the present results support the idea that it is the stripe order and not the LTT distortion which is responsible for the suppression of the 3D coherent superconducting state.

Taking this one step further, these results are also consistent with the scenario of a system which, in the presence of stripe order, is a 2D superconductor [8]. In this case, the superconductivity is proposed to be in the form of a "pair density wave" state [9], where the superconducting order parameter is modulated by the periodic potential resulting from the charge order, with twice the period. This, and the 90° rotation of the charge stripes along the c-axis, provides a natural explanation for the destructive interference of the Josephson currents. In this scenario, when the mid-IR pulse melts the charge order, it removes the periodic potential. At that point, the superconducting condensates are free to coherently couple along the c-axis and do so on a time scale of the Josephson plasma resonance, i.e. a few hundred femtoseconds.

It is worth noting that such laser-induced transitions into non-thermal electronic or structural phases is a rather general phenomenon, as observed in other condensed matter systems [15,25,26,27]. The mechanism by which the charge order is melted following the excitation of the Cu-O bond stretching mode in the present case is not yet clear. Further microscopic modeling of the charge order and the LTT structure in the presence of a strongly driven lattice phonon [28,29] could aid understanding this physical pathway. In addition, one might take into account the possibility of direct excitation of an electronic mode that is associated with the pair density wave state in LBCO$_{1/8}$ as described in Ref. 23. A direct coupling of the mid-infrared pulse to such an electronic excitation could explain the fast melting of the charge stripe order, and the weak effect on the lattice distortion.

## ACKNOWLEDGEMENTS

The authors thank J.M. Tranquada for helpful discussions.

Portions of this research were carried out on the SXR Instrument at the Linac Coherent Light Source (LCLS), a division of SLAC National Accelerator Laboratory and an Office of Science user facility operated by Stanford University for the U.S. Department of Energy. The SXR Instrument is funded by a consortium whose membership includes the LCLS, Stanford University through the Stanford Institute for Materials Energy Sciences (SIMES), Lawrence Berkeley National Laboratory (LBNL, contract No. DE-AC02-05CH11231), University of Hamburg through the BMBF priority program FSP 301, and the Center for Free Electron Laser Science (CFEL). Work performed at Stanford University was further supported from U. S. Department of Energy, Office of Basic Energy Science, Division of Materials Science and Engineering under the contract no. DE-AC02-76SF00515.

Work performed at Brookhaven was supported by US Department of Energy, Division of Materials Science under contract no. DE-AC02-98CH10886. The research leading to these results has received funding from the European Research Council under the European Union's Seventh Framework Programme (FP7/2007-2013) / ERC Grant Agreement n° 319286 (Q-MAC), and from the German Research Foundation (DFG-SFB 925).

# FIGURE CAPTIONS

**Fig. 1:** Schematic drawing of the $La_{1.875}Ba_{0.125}CuO_4$ charge, spin and lattice arrangement within a $CuO_2$ plane in the stripe-ordered, low-temperature tetragonal phase (T<55 K). Here, Cu atoms are shown as blue, oxygen atoms as red spheres. Holes form stripes which separate domains of oppositely phased antiferromagnetic domains (spins indicated by arrows). The LTT distortion is visible through the tilt of the octahedra central planes. The periodic stacking of those $CuO_2$ planes is sketched in the lower figure part. The stripe orientation rotates by 90 degrees between layers.

**Fig. 2:** (a) Top view of the experimental setup, shown for the diffraction condition of the charge stripe order peak (see text for details). (b) Transient intensity of the charge stripe order diffraction peak in (001) cleaved $La_{1.875}Ba_{0.125}CuO_4$ measured at the (0.24 0 0.5) wave vector. Resonant mid-IR excitation with 1.9 mJ/cm$^2$ fluence at zero time delay results in a prompt decrease of the scattered intensity on the sub-ps time scale. The red solid line represents an exponential function with a time constant set to 300 fs, i.e. the resolution of the experiment. The upper panel shows the diffracted spots, recorded with the CCD camera and averaged over about 20,000 FEL shots, at negative and positive time delays.

**Fig. 3:** Light-induced changes in the intensity of the (001) diffraction peak reflecting the LTT distortion. Again, the $La_{1.875}Ba_{0.125}CuO_4$ crystal is excited with 1.9 mJ/cm$^2$ fluence of mid-infrared light. The red solid is a single exponential fit to the data yielding a time constant of 15 ps. The inset shows the same data set with an expanded y-axis. The diffraction spots recorded on the CCD, averaged over 400 FEL shots for a positive and a negative time delay, are shown in the upper panel.

**FIGURES**

**Fig. 1**

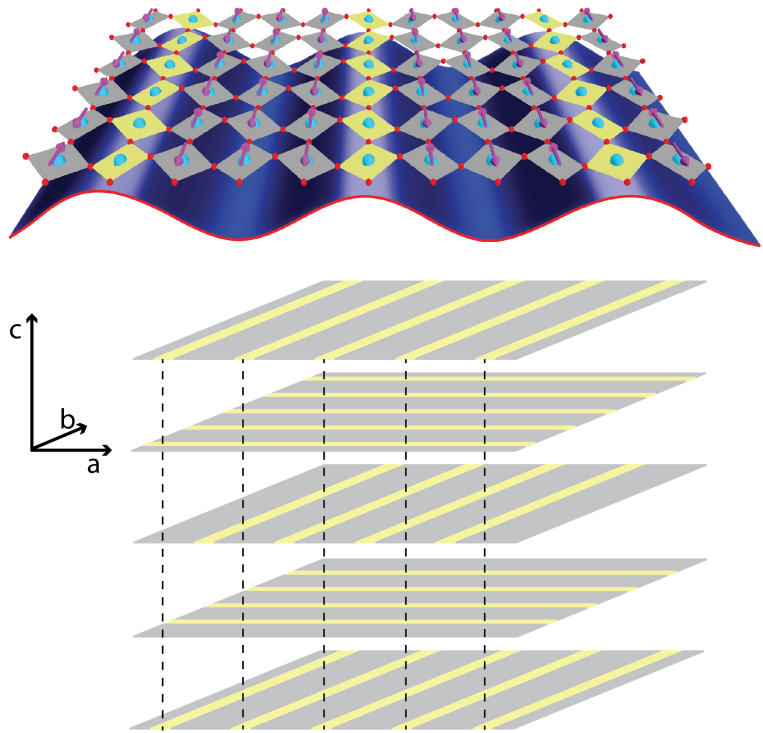

**Fig.2**

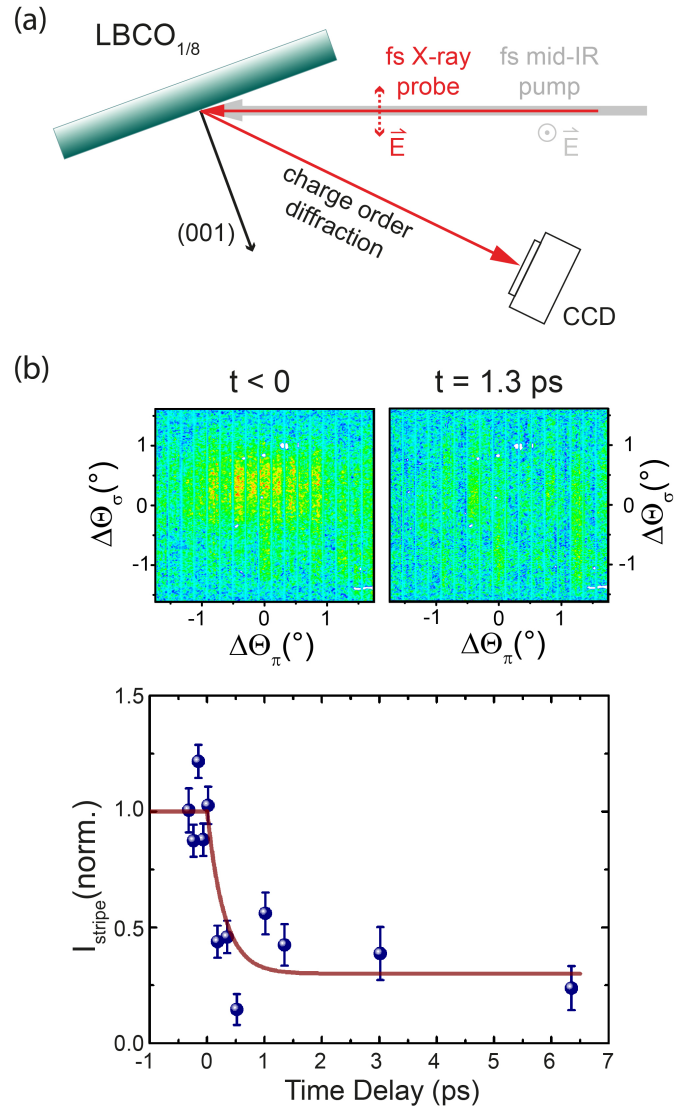

**Fig.3**

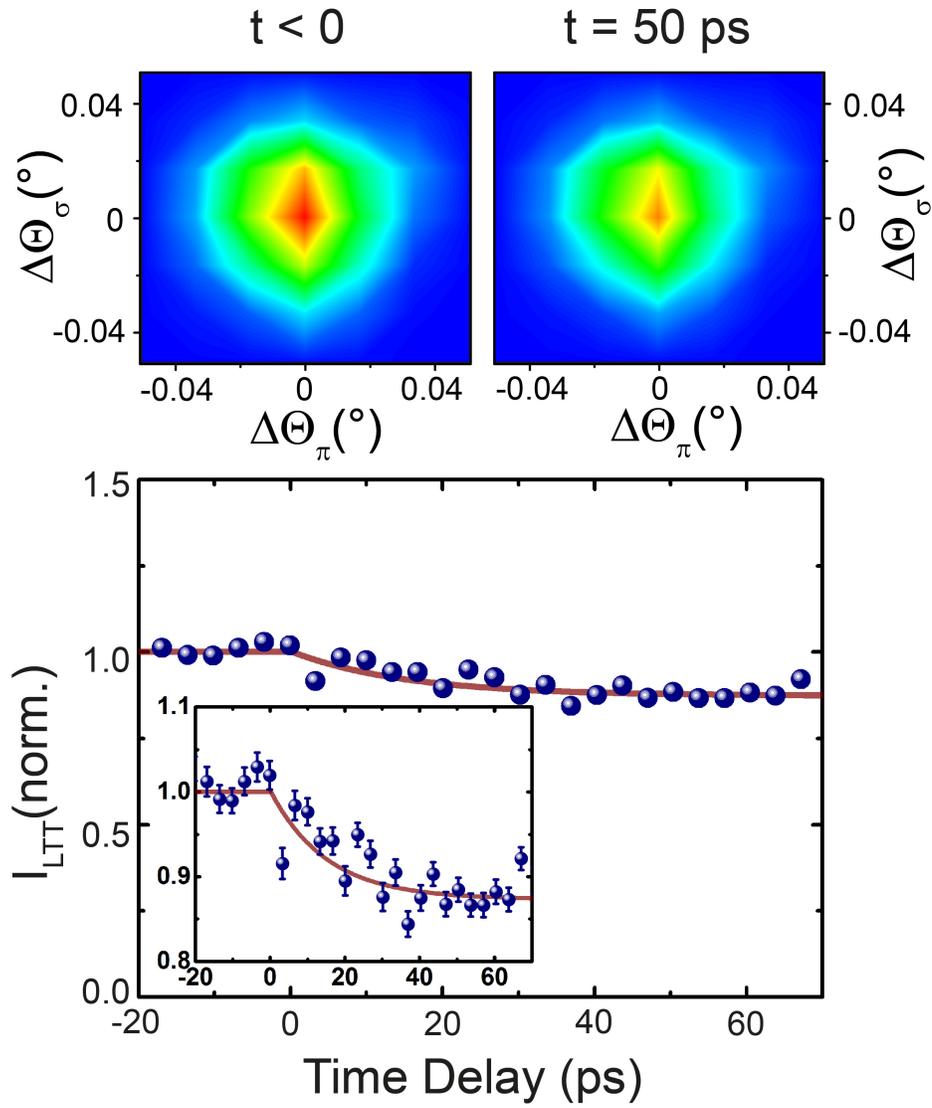

# REFERENCES

\* michael.foerst@mpsd.mpg.de

† andrea.cavalleri@mpsd.mpg.de

§ hill@bnl.gov